\documentstyle{mn}

\newcommand{\ale}{\ \raisebox{-.3ex}{$\stackrel{<}{\scriptstyle \sim}$}\ }
\newcommand{\age}{\ \raisebox{-.3ex}{$\stackrel{>}{\scriptstyle \sim}$}\ }
\newcommand{\omb}{\Omega_{\rm orb}}
\newcommand{\Pb}{P_{\rm orb}}
\newcommand{\vsc} {$d\,\Omega_{\rm orb}$}
\newcommand{\lsc} {$d$}
\newcommand{\tsc} {$\Omega_{\rm orb}^{-1}$}
\newcommand{\lunit} {$M\,d^2\,\Omega_{\rm orb}^{3}$}
\newcommand{\msol}{{\rm M_\odot}}

\input psfig

\title[Superhumps in the VY Scl stars]{Eccentric discs in binaries
with intermediate mass ratios:\\ Superhumps in the VY Sculptoris stars}
	
\author[J.R. Murray, B. Warner \& D.T Wickramasinghe]
{J.R. Murray$^1$\thanks{Present address: Department of Physics \&
Astronomy, University of Leicester, University Road, 
Leicester LE1 7RH, UK}, B. Warner$^2$ and
	D.T Wickramasinghe$^1$\\
	$^1$ The Astrophysical Theory Centre, 
	Australian National University, ACT 0200, Australia \\
 	$^2$ Department of Astronomy, University of Cape Town, 
        Rondebosch 7700, South Africa}	
 	
\begin{document}

\maketitle

\begin{abstract} 
We investigate the role of the eccentric disc
resonance in systems with mass ratios $q \age 1/4$, and demonstrate the effects
that changes in the mass flux from the secondary star have upon the disc
radius and structure. 
The addition of material with low specific angular momentum to 
its outer edge restricts a disc
radially. Should the mass flux from the
secondary be reduced, it is possible for the disc in a system with
mass ratio as large as $1/3$ to expand to the $3:1$ eccentric inner 
Lindblad  resonance and for superhumps to be excited.
\end{abstract}


\section{Superhumps in Nova-like Variables}
The superhump phenomenon was first observed in superoutbursts of dwarf
novae of short orbital period ($P_{\rm orb} \ale 2.8\,{\rm h}$: see
Warner 1995b). It can be explained as an effect of an eccentric disc
that precesses with respect to the tidal field of the secondary star
(Whitehurst 1988). 
In low mass ratio binaries ($q \ale 1/4$), disc
eccentricity can be excited via tidal resonance (Whitehurst 1988,
Hirose \& Osaki 1990, Lubow 1991). In systems with larger $q$
however, the disc is truncated at too small a radius for the resonance
to be effective.

Observationally there is evidence that discs can be tidally resonant
for $q$ significantly greater than $1/4$.  Superhumps appear in high
mass transfer discs (those of nova-like variables and of nova
remnants) in some systems with orbital periods up to 3.5 h (Skillman
et al. 1998). At this period the mass of the secondary star is
$0.31\,\msol$ (equation 2.100 of Warner 1995b). Unfortunately there
are no reliable determinations of $q$ in the superhumping systems with
$\Pb > 3$h. The mean mass of the primaries of the nova-likes is $\sim
0.74\,\msol$ (Warner 1995b) and the range is likely to be at least
$0.5 - 1.0\,\msol$. If the primary mass is at the upper end of this
range, 
it is therefore possible that a system with $\Pb \sim 3.5$h would
have a mass ratio approaching that needed for the eccentric resonance
to lie within the accretion disc.

There can be little doubt that a mechanism is required to extend the
range of mass ratios for which tidal resonance occurs,  
beyond the $q \ale 1/4$ established in
equilibrium disc calculations. One possibility is suggested by
the VY~Scl stars, systems  with orbital periods in the range
$3.0 < \Pb < 4.0$ h (Warner 1995b) in which mass transfer is
liable to reduce significantly  at irregular intervals. 
Such a drop in $\dot M_{\rm s}$ would allow a tidally stable,
equilibrium disc that just
marginally fails to reach the eccentric resonance to expand radially
and become eccentrically unstable.

In this paper we demonstrate that the discs in high $\dot M$ systems
can be
excited into an eccentric state when the mass transfer
from the secondary drops below its equilibrium value. In the next
section we discuss the observational peculiarities of the VY~Scl
stars. In section three we outline our model for superhumps in these
systems. We present numerical results that support our case in section
four, and make our conclusions in section five.

\section{Mass Transfer in Nova-like Variables}
The long-term light curves of nova-like variables show brightness
variations on a range of time-scales. Quasi-periodic modulations with
ranges $\sim 0.7$ mag and time-scales of $1-2$ months may be intrinsic
to the disc and in effect suppress dwarf nova outbursts in which the
thermal instability of the disc is modified by irradiation from the
white dwarf primary (Warner 1995a). Larger brightness variations on
time-scales from days to years are also seen, in which the star shows
a ``bright state'' before falling from 1 to 5 magnitudes in
brightness. This behaviour is named after the type star, VY~Sculptoris, and
is generally believed to be the result of significant and relatively
rapid drops in mass transfer from the secondary.

The known VY~Scl stars (Table 4.1 in Warner 1995b) have orbital
periods in the range $3.2 \leq \Pb \leq 4.0$ h. Two apparently
discrepant systems, V794~Aql and V745~Cyg have since had their orbits
redetermined. Thorstensen (private communication) found the orbital
period of V745 Cyg to be $4.05$ h, and Honeycutt \& Robertson (1998)
found the period of V794~Aql to be $3.68$ h.
Furthermore, every nova-like
variable in the 3.0 to 4.0 h $\Pb$ range that has a well observed
long-term light curve has been found to be of VY~Scl type (Warner 1995b).
Unstable mass transfer in these systems may be due to irradiation of
the secondary by the primary (Wu, Wickramasinghe \& Warner, 1995).

As a consequence, in the range $3.0 \ale \Pb \ale 4.0$ h, high $\dot
M$, stable accretion discs are subject to occasional reductions of
received mass at their outer edge. In the following sections we
numerically investigate the disc response.

\section{Tidal instability in systems with intermediate mass ratios}
Eccentricity can be excited in a close binary accretion disc if it overlaps or
very nearly overlaps the $3:1$ eccentric inner Lindblad resonance
(Lubow 1991). With a characteristic radius almost half the
interstellar separation, 
this resonance is only accessible to discs in low mass ratio systems.
Tidal forces prevent 
discs in high mass ratio systems from reaching the
resonance. 

Paczy\'nski (1977) approximated the streamlines in a pressureless, inviscid
disc with  ballistic particle trajectories. 
He then suggested that such a disc could not be larger than the
largest ballistic orbit that doesn't intersect orbits interior to it.
Using Paczy\'nski's estimate, Hirose \& Osaki (1990), and Lubow
(1991) found that the $3:1$ resonance could only be reached if $q
\ale 1/4$.
 
But the tides that truncate a close binary disc also remove angular
momentum from it. 
Papaloizou \& Pringle (1977) calculated the tidal removal of 
angular momentum
from a disc comprised of gas in linearly perturbed circular orbits. 
They estimated that the 
disc would need to extend out to
 approximately $0.85-0.9$ times the mean Roche lobe radius for tidal
torques to be effective (slightly beyond the Paczy\'nski
radius). However the assumption of linearity made by Papaloizou \&
Pringle lead to an underestimate of the dissipation in the
neighbourhood of Paczy\'nski's orbit crossing radius and thus to a
larger estimate for the tidal radius.
 Thus when Whitehurst \&
King (1991) used the Papaloizou \& Pringle value for maximum disc
size, they estimated that  the $3:1$ resonance could be
reached in systems with $q$ as high as $1/3$.

For near equilibrium discs subjected to a fixed rate of mass addition from
the secondary, the numerical results are more in keeping with the lower
mass ratio limit of $1/4$. Hirose \& Osaki (1990) obtained eccentric discs in
simulations with $0.10 \leq q \leq 0.20$, and Murray (1998) for mass
ratios $\leq 0.25$. Whitehurst (1994) found eccentricity in
simulations at mass ratios as high as $1/3$. However his discs were initialised
 with a rapid burst of mass transfer and so were not close to an
equilibrium when they encountered the $3:1$ resonance. 

As discussed in the previous section, the observations are not
consistent with a mass ratio limit as low as $1/4$. We propose then
that higher mass ratio systems can become resonant during periods when
the disc is not in equilibrium with the mass transfer stream.  

Gas passing through  the inner Lagrangian point has a specific
angular momentum with respect to the white dwarf equivalent to that of
a circular Keplerian orbit of radius  $r_{\rm circ}$. This {\em
circularisation radius} is of the order of a tenth to one fifth 
the interstellar
separation and is very much inside the disc's tidal truncation limit
(Lubow \& Shu, 1975).
Hence, mixing with the
mass transfer stream reduces specific angular momentum in the
outer disc and restricts the disc radially. Of course 
the principal
confinement of the disc is provided by tidal torques, 
however the additional
confinement due to the stream can be significant.
If, as is suspected to
happen in the VY~Scl stars, $\dot M_{\rm s}$ is suddenly reduced, then
the radial restriction on the disc will be  relaxed,
and the disc will expand.
The  expansion can be only temporary, as the
disc will subsequently adjust to the new  
$\dot M_{\rm s}$ on its viscous time-scale, and return to its initial
size. However, we will show in the next section that 
the period of readjustment can be sufficient to allow
the disc to encounter the $3:1$ resonance and become eccentric.

\section{Results}
\subsection{Numerical Method}
In this section we present numerical results obtained using a two
dimensional, fixed spatial resolution, smooth particle hydrodynamics
(SPH) code. Previous calculations and detailed descriptions of the
code are to be found in Murray (1996, 1998). Although the code has
been improved to allow for variable spatial resolution (see
e.g. Murray, de~Kool \& Li, 1998) and for the three dimensional
modelling of accretion discs (Murray \& Armitage, 1998), we found that
by restricting ourselves to two dimensions and fixed resolution, we
could more thoroughly
explore parameter space with a reasonable expenditure of computer
time. Our ability to capture the essential features of the eccentric
instability is not compromised, and previous results have been
consistent with those of three dimensional calculations 
(see e.g. Simpson \& Wood, 1998).

The simulations described here demonstrate that, in systems with
intermediate mass ratio, a sudden reduction in mass
transfer from the secondary star can allow an equilibrium disc to
expand so as to 
come into contact with the eccentric resonance. 

Two sequences of simulations were completed. In the first series  we subjected
near-equilibrium discs to sudden reductions in $\dot M_{\rm s}$, the
aim being to demonstrate the viability of the proposed mechanism. 

We then completed a second set of  simulations in which we replaced
the mass transfer stream with steady mass
addition  at the circularisation radius, $r_{\rm circ}$. 
By removing the additional constraining effect of the stream, and
allowing the disc to reach a {\em fully extended} steady state,  we 
made it as easy as possible for the disc to interact with the
eccentric resonance. We ran such simulations for several values of $q$ and
thus determined the range of mass ratios for which a reduction in mass
transfer might result in superhumps.

As in Murray (1996) and Murray (1998), we use the interstellar
separation $d$, the binary mass $M$, and the inverse of the orbital
angular velocity \tsc, as the length, mass and time scalings for our
simulations.  

\begin{figure}
\psfig{file=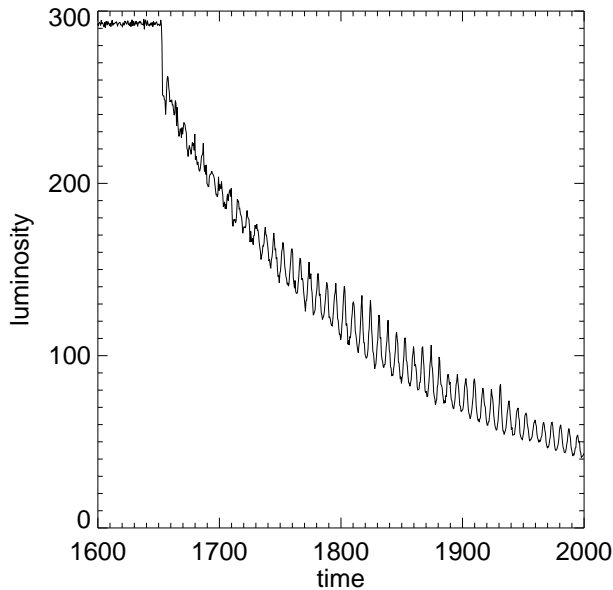,width=8cm}
\caption{Energy dissipation ``light curve'' from the outer disc ($r>0.2$) of the
$q=0.29$ simulation with mass transfer from the secondary  shut off at
time $t=1652\,$\tsc. The contribution of the inner disc to the
dissipation was not included for clarity. The units for the luminosity
are \lunit.}
\label{VYScldiss}
\end{figure}

\begin{figure}
\psfig{file=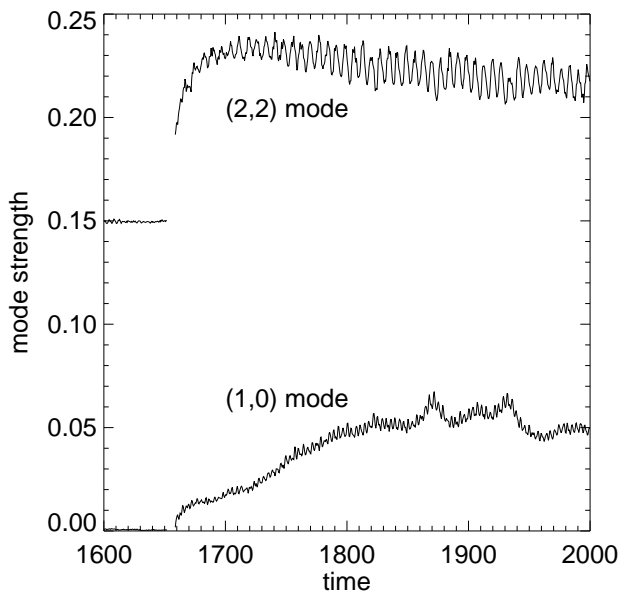,width=8cm}
\caption{The  response of the disc eccentricity and the $l=m=2$ tidal
mode to a complete cessation of mass transfer from the secondary.
See figure~\ref{VYScldiss} for the corresponding light curve.}
\label{VYSclmodes}
\end{figure}

\subsection{Variable $\dot M_{\rm s}$ Calculations}
In Murray (1998) we described a    calculation for mass ratio $q=0.29$
that failed to excite the eccentric resonance (simulation 1 of that
paper). In that calculation mass was introduced at a constant rate of one
particle per $\Delta\,t=0.01\,$\tsc\, at the inner Lagrangian
point. An isothermal equation of state was used with
$c=0.02\,$\vsc. The spatial resolution was fixed  with the smoothing
length $h=0.01\,$\lsc. The high rate of viscous dissipation that
occurs in CV discs in outburst was modelled using an artificial
viscosity term (Murray 1996, 1998). The artificial viscosity parameter
$\zeta=10$. The gave an effective Shakura-Sunyaev parameter $\alpha
\simeq 1.9$ at the resonance radius. This was somewhat on the high side
but made the calculations more manageable.
At the end of the simulation ($t=1652.00\,$\tsc) the non-resonant disc
had  an equilibrium mass of 
19950 particles. 

\begin{figure*}
\psfig{file=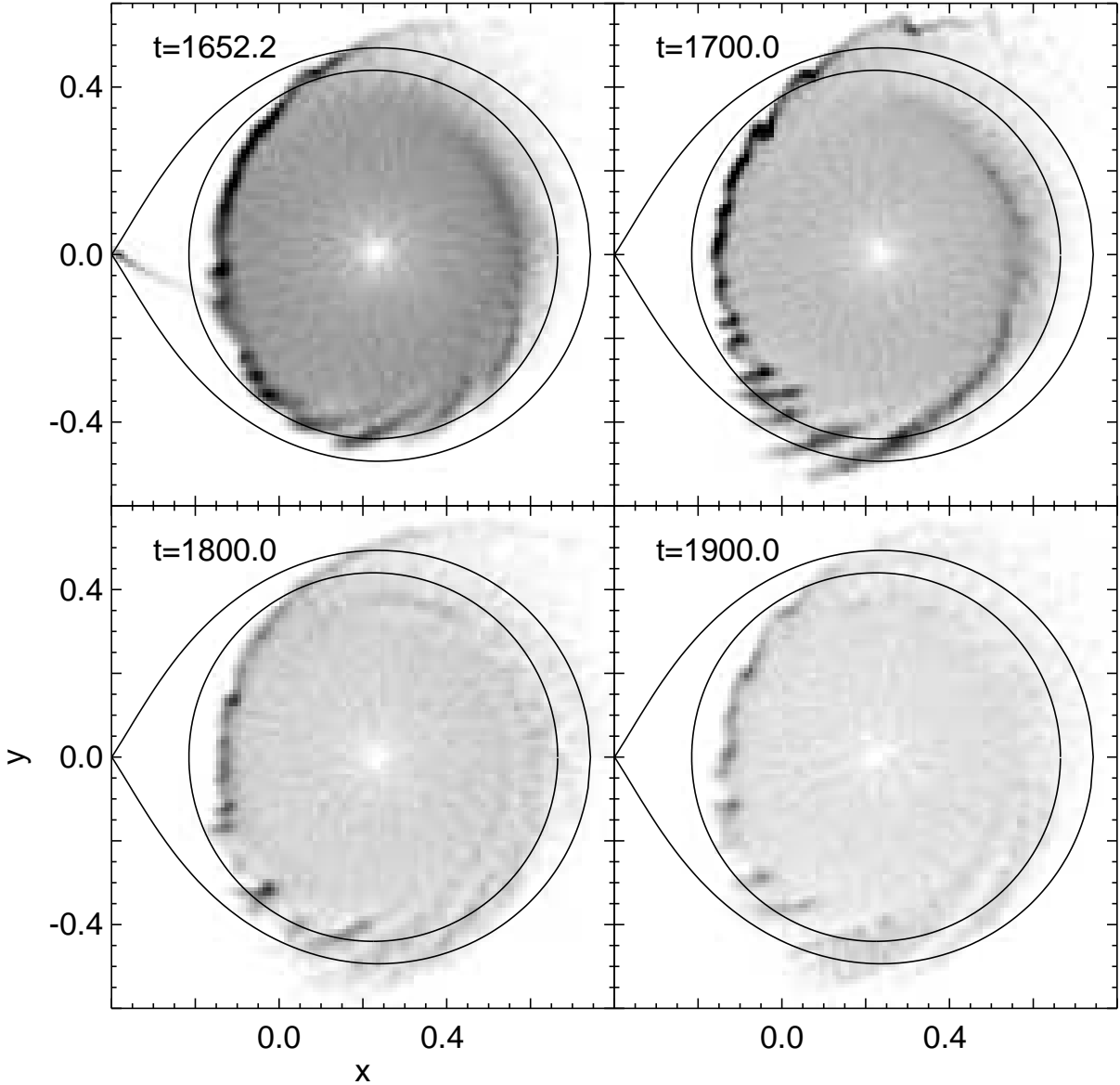,width=16cm}
\caption{Grey scale density maps for the first $q=0.29$ simulation, covering
the time interval shown in figures~\ref{VYScldiss}
and~\ref{VYSclmodes}. The same absolute density scale is used in each
frame. $\dot M_{\rm s}$ was set to zero at $t=1652.00\,$\tsc. The
Roche lobe of the primary and the $3:1$ resonance are shown as solid lines.}
\label{VYScldensity}
\end{figure*}

For this paper we used the final
state of the above calculation as the {\em initial state} for a new simulation
with $\dot M_{\rm s}~=~0$ (no mass addition). All other system parameters
were left  unchanged.
The disc evolution was then followed to
$t=2000.00\,$\tsc, by which time there were only 5031 particles
remaining in the disc. By plotting the energy viscously dissipated in
the disc as a function of time we obtained a ``light curve'' for the
simulation (figure~\ref{VYScldiss}).  The
stream-disc impact is responsible for a significant fraction of the
energy dissipation in the outer disc. Hence when the accretion
stream was shut off (at $t=1652.00\,$\tsc) there was a sudden drop in
disc luminosity. Almost immediately after mass transfer ceased the
accretion disc became resonant and eccentric. The subsequent motion of the
eccentric mode {\em relative to the binary frame} introduced a cyclic
contribution to the energy dissipation in the outer disc (this
is discussed in more depth in Murray, 1998) which we identify as a
superhump signal. Figure~\ref{VYScldiss} shows the superhumps reaching
maximum amplitude somewhere between $t=1800$ and $1900\,$\tsc. The
subsequent decay of both the superhump amplitude and of the background
luminosity was simply due to the decline in disc mass. 

As the simulation ran we Fourier analysed the disc's density
distribution using the technique described in Lubow (1991). 
Each Fourier mode's variation in azimuth and time is given by
$l\,\theta-m\,\omb\, t$ where $\theta$ is azimuth 
measured in the inertial frame,
$\omb$ is the angular velocity of the binary, and $l$ and $m$ are
integers.
Hence the disc's eccentricity is measured by  the
$l=1$, $m=0$ mode, and the largest amplitude non-resonant tidal
response has $l=m=2$. In some sense this latter mode (which appears in
density maps as a two armed spiral pattern fixed in the binary frame)
indicates the radial extent of the disc. The larger the tidal
response, the more disc material there is at large radii.

The strengths of the eccentric and spiral modes are plotted in 
figure~\ref{VYSclmodes}. The tidal mode strength jumped sharply once
mass addition ceased, strongly indicating a redistribution of matter
radially outward through the disc. We confirmed this interpretation 
by checking the actual particle distributions. 
Immediately prior to the demise of the mass transfer stream, $18.2\,\%$
of the disc mass was at radii $r>0.4\,$\lsc. By $t=1700.00\,$\tsc, 
this figure had increased to $25.6$\,\%. There was also an increase in absolute
terms  from  3636 to 4046 particles.

The eccentric mode also became immediately stronger upon cessation of
mass transfer, and then continued to grow as the resonance took effect.
The maximum in the eccentric mode strength coincided with the largest
amplitude superhumps.

We have produced a sequence of grey scale density maps of the disc 
(figure~\ref{VYScldensity}), from which it is clear that the mass transfer
stream did indeed restrict the disc's radial extent. In its original
equilibrium state (top left panel), 
the disc immediately downstream of
the stream impact region lay entirely within the Roche lobe of the
primary. 
Compare that with the disc shortly after mass transfer had ceased (top
right) which extended beyond the Roche lobe.

\begin{figure}
\psfig{file=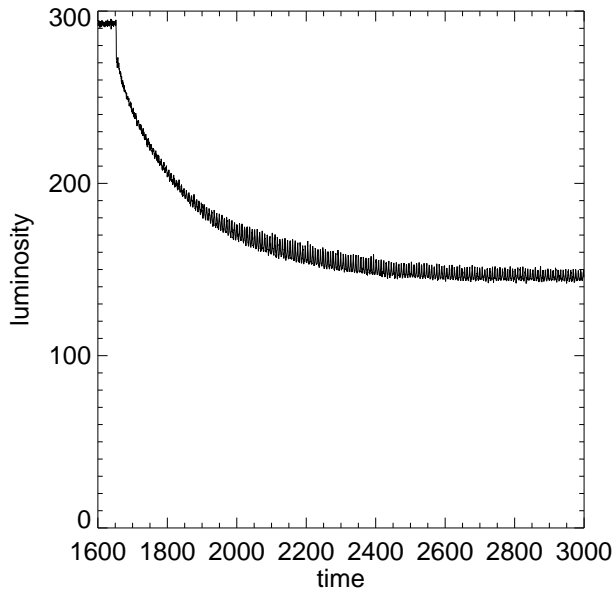,width=8cm}
\caption{Energy dissipation ``light curve'' from the outer disc ($r>0.2$) of the
$q=0.29$ simulation with mass flux from the secondary halved
at time $t=1652\,$\tsc. The units for the luminosity
are \lunit. }
\label{VYScldiss2}
\end{figure}

\begin{figure}
\psfig{file=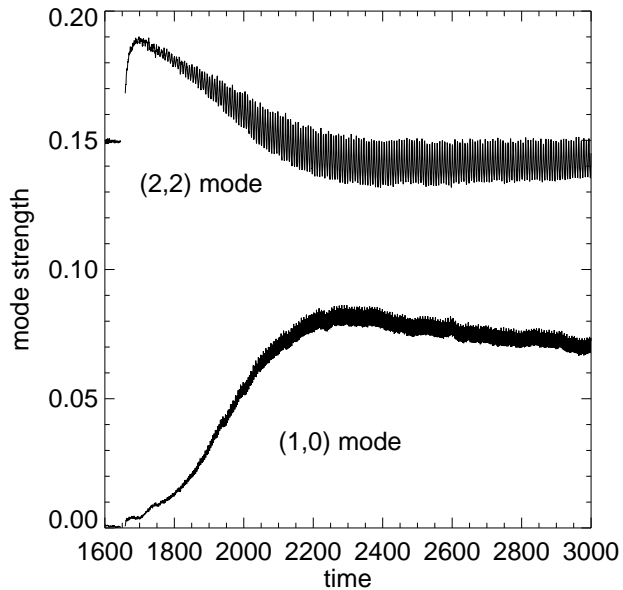,width=8cm}
\caption{Eccentric and tidal mode strengths corresponding to
figure~\ref{VYScldiss2} light curve. 
The mass flux from the secondary was halved
at time $t=1652\,$\tsc.}
\label{VYSclmodes2}
\end{figure}

Having determined  the mechanism to be viable, we then investigated
whether disc eccentricity could be excited with a less drastic
reduction in $\dot M_{\rm s}$. We completed a second calculation with
identical initial conditions, but with the mass
transfer rate reduced to half its initial value instead of being extinguished
completely.  
Figures~\ref{VYScldiss2} and
\ref{VYSclmodes2} follow the evolution of the disc to $t=3000$\,\tsc,
by which time the disc mass had stabilised at approximately 10500
particles.

As in the first simulation, the disc immediately came into contact
with the resonance and became more eccentric. This time however the
superhump amplitude was much reduced.
The eccentricity reached a maximum at $t \simeq 2300$\,\tsc\, and then
declined. Note that despite the reduction in superhump amplitude, the
amplitude of the $(1,0)$ mode was larger in the second calculation (compare
figures \ref{VYSclmodes} and \ref{VYSclmodes2}). Whereas the eccentric
mode strength is a measure of the eccentricity as a whole, it does not
directly measure the eccentricity of individual particle orbits. The
superhump signal, however, is generated by those particles on the 
most eccentric
orbits. In the second simulation, the continued mass transfer 
restricted the maximum eccentricity which could be acquired by gas in
the outer disc, and hence limited the superhump amplitude.

We had expected {\it a priori}  that the disc would return to an axisymmetric
state in equilibrium with
the reduced mass flux from the secondary. However
we followed this calculation to $t=5700$\,\tsc\, and were
surprised to find that the eccentricity did not decline to zero as
expected, but
instead stabilised with $0.05 < S_{(1,0)} < 0.07$. The plot of the
tidal $(2,2)$ mode (figure~\ref{VYSclmodes2}) and 
density maps (not shown) indicate that the disc returned to its original
radial extent by time $t \simeq 2100$\,\tsc.
There is thus the intriguing possibility that this eccentric state,
once reached, was quasi-stable.  

\begin{figure}
\psfig{file=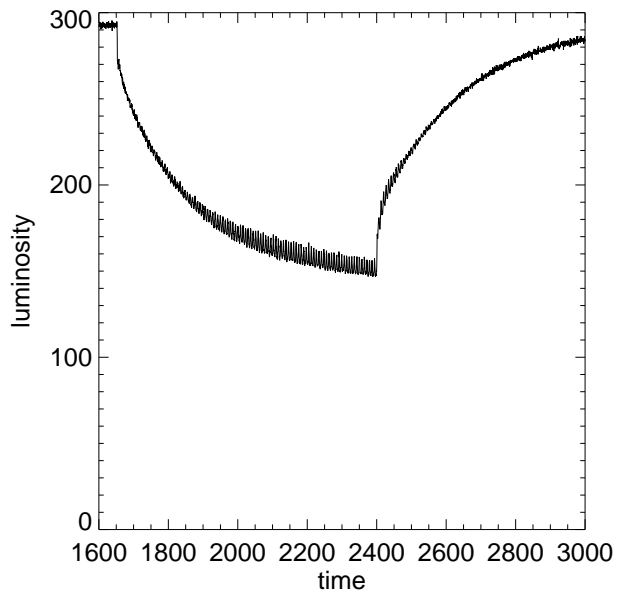,width=8cm}
\caption{Energy dissipation ``light curve'' from the outer 
disc ($r>0.2$) of the
$q=0.29$ simulation with mass flux from the secondary halved
at time $t=1652\,$\tsc, and then restored to its original value at
$t=2400\,$\tsc. The units for the luminosity
are \lunit.}
\label{VYScldiss3}
\end{figure}

\begin{figure}
\psfig{file=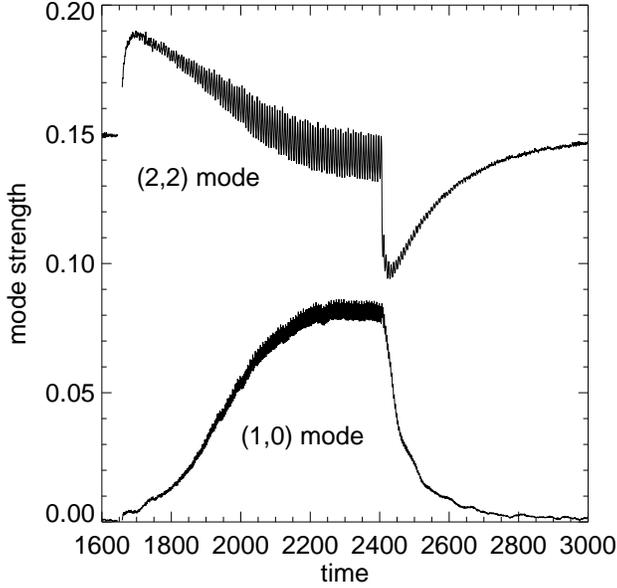,width=8cm}
\caption{Eccentric and tidal mode strengths corresponding to
figure~\ref{VYScldiss3} light curve. 
The mass flux from the secondary was halved
at time $t=1652\,$\tsc, and then restored to its original value at $t=2400\,$\tsc.}
\label{VYSclmodes3}
\end{figure}

To complete this section we performed a third simulation, in which we
took the
previous simulation at time $t=2400.00$\,\tsc\,
and restored the mass transfer rate to its original
value. Figures~\ref{VYScldiss3} and \ref{VYSclmodes3} show that the
disc was initially  forced rapidly inwards and circularised by the
increased mass flux from the secondary. Further adjustment then
occurred on the viscous time-scale as the disc re-expanded to its
equilibrium radius and mass flux, and the eccentricity decayed
away. By the end of the calculation ($t=2400.00$\,\tsc) the disc had
returned to its original axisymmetric state. 

\subsection{Stable $\dot M_{\rm s}$ Calculations}
In this section we describe  four simulations of discs built
up to steady state from zero initial mass 
via {\em steady} mass addition at the
circularisation radius. Although  one
would be hard-pressed to find such a disc in nature,
it is perhaps the most favourable disc configuration for the
excitation of the eccentric resonance.
Whereas in the simulations of the previous section the interaction
with the resonance was limited to the viscous time-scale, there is no
such time limitation in these calculations. 
Thus if, for a given mass ratio,  
eccentricity cannot be excited with this artificially
favourable setup, then it is hard to see how it could be excited with
more realistic mass input. Hence we can obtain a limiting mass ratio
for which mass transfer reductions can inspire superhumps. 

As in the previous section we took an
isothermal equation of state with $c=0.02\,$\vsc, an SPH
smoothing length $h=0.01\,$\lsc, and set the SPH artificial viscosity
parameter $\zeta=10$. Each calculation was terminated at $t=1000$\,\tsc.

\begin{table}
\caption{Summary of steady $\dot M_{\rm s}$ simulations}
\label{tab:rcirc}
\begin{tabular}{ccc}
$q$ & final $S_{(1,0)}$ & ${P_{\rm d}}/{P_{\rm orb}}$\\
\hline
$0.29$ & $4.5 \times 10^{-2}$ & $1.153 \pm 0.008$\\
$0.30$ & $3.5 \times 10^{-2}$ & $1.163 \pm 0.005$\\
$0.31$ & $5.9 \times 10^{-3}$ & $1.173 \pm 0.007$\\
$1/3$  & $2.7 \times 10^{-3}$ & --\\
$0.29^\dagger$ & $6.3 \times 10^{-2}$ & $1.1583 \pm 0.0001$\\
\hline
\end{tabular}
\\
${}^\dagger$ simulation from section 4.2 with $\dot M_{\rm s}$ half
initial value, added at $L_1$.
\end{table}
 Simulations were completed for $q=0.29,0.30,0.31$ and $1/3$. In each
case, mass was added at a fixed rate of
 one particle per
$\Delta\,t=0.01\,$\tsc, at a radius
$r=0.1781\,$\lsc\,   ($r_{\rm circ}$ for a system with $q=0.18$). 
By doing this we have kept the  specific angular
momentum of newly added material constant across the simulations. 
However the angular momentum is somewhat greater 
(14\% in the case of $q=1/3$)
than if we had used the correct circularisation radius for each mass
ratio.

The strength of
the eccentric mode at the conclusion of each simulation, and the disc
precession periods are listed in table~\ref{tab:rcirc}. The higher the
mass ratio, the more weakly the eccentric resonance was excited.
For the
$q=1/3$ simulation  some volatility could still 
be discerned in the eccentric mode
strength but no superhumps were apparent either in the light curve or
in its power spectrum. 

We included in table~\ref{tab:rcirc}  the second of the
simulations completed for section 4.2 (for which $\dot M_{\rm s}$ was
reduced a factor 2). Given its length (four times
that of the calculations completed for this section), this calculation
provided a very reliable  measurement
of $P_{\rm d}$ and thus a useful check on the other results tabulated here.

A mass ratio of approximately $1/3$ would appear to be the upper limit
beyond which the eccentric resonance cannot be excited even temporarily.
Our results correspond well with those of Whitehurst (1994) who found
superhumps in a simulation with $q=1/3$ but not in a simulation with
$q=0.34$. He constructed his discs with a massive initial mass
transfer burst, followed by a much reduced but constant $\dot M_{\rm s}$.
The mass distribution in Whitehurst's discs would have been therefore  more
akin to a disc built up via mass addition at $r_{\rm circ}$ than to a
disc subject to steady mass addition from $L_1$. 

\section{Conclusions}
Under conditions of
steady mass transfer, a mass ratio $\ale 1/4$ is required for
a close binary accretion disc to encounter the $3:1$ eccentric inner
Lindblad resonance. 
However, it is possible for eccentricity to be excited in the disc of a high
mass transfer system with $q \ale 1/3$ if $\dot M_{\rm s}$ is reduced,
as is thought to occur in the VY~Sculptoris systems. 
Our simulations suggest that
the disc will remain eccentric as long as  $\dot M_{\rm s}$ 
remains at the lower value. 

Thus, a precessing, eccentric disc remains the best explanation of
superhumps, even for systems with $3.0 < \Pb < 4.0$ h.

\end{document}